\documentclass[aps,pra,twocolumn, amsmath, amssymb,superscriptaddress]{revtex4}
\usepackage{graphicx}
\usepackage{dcolumn}
\usepackage{bm}
\usepackage{indentfirst}
\usepackage{xcolor}

\begin{document}

\title{Quantum theory of magneto-optical trap}

\author{O.N. Prudnikov}
\email{oleg.nsu@gmail.com}
\affiliation{Institute of Laser Physics, 630090, Novosibirsk, Russia}

\author{A. V. Taichenachev}
\affiliation{Institute of Laser Physics, 630090, Novosibirsk, Russia}
\affiliation{Novosibirsk State University, 630090, Novosibirsk, Russia}

\author{V. I. Yudin}
\affiliation{Institute of Laser Physics, 630090, Novosibirsk, Russia}
\affiliation{Novosibirsk State University, 630090, Novosibirsk, Russia}
\affiliation{Novosibirsk State Technical University, 630073, Novosibirsk,
Russia}

\author{L. Zhou}
\affiliation{Innovation Academy for Precision Measurement Science and Technology, Chinese Academy of Sciences, Wuhan 430071, China}

\author{M.S. Zhan}
\affiliation{Innovation Academy for Precision Measurement Science and Technology, Chinese Academy of Sciences, Wuhan 430071, China}

\date{\today}

\begin{abstract}
We present a quantum theory of a magneto-optical trap (MOT) from first principles based on the quantum kinetic equation for the atomic density matrix with taking into account the recoil effects caused by the interaction of atoms with the laser field. An efficient method for solving the quantum kinetic equation is proposed. It is shown that the steady-state solution describing the atoms in the MOT has a significantly non-equilibrium nature and can be described within the framework of a two-temperature distribution. The momentum distribution of cold atoms in the MOT depends on the magnetic field gradient and, in general, significantly differs from the momentum distribution of atoms in the optical molasses, which is usually used as an approximation to describe the MOT. We have also shown that with an increase in the magnetic field gradient, a spatial two-component distribution of atoms in the trap is formed even for a single particle approximation when interatomic interactions are neglected.
\end{abstract}


\maketitle


\section{Introduction}
Magneto-optical trap (MOT) is currently one of the effective and robust tool for trapping and laser cooling of neutral atoms \cite{raab1987,metcalf}. The MOT operation is based on the combination of laser cooling, caused by dissipative light pressure forces, and a deep potential (of the order of several $\mbox{K}$), created by these forces in a spatially non-uniform magnetic field. Cold atoms prepared in MOT have wide applications in various areas \cite{Katori2020,Lion2017,Ludlow2018,Nicholson,BERMAN1997xiii}, such as high-resolution spectroscopy, studying cooling and dynamics of atoms in optical lattices, for precise frequency standards, atom interferometry, and so on.

The theory of MOT is usually described within the semiclassical approach in terms of dissipative light forces on atoms \cite{raab1987, metcalf}. The temperature limit of laser cooling in MOT, as well as in optical molasses, is determined by the balance between the processes of kinetic energy dissipation, caused by the forces on atoms from counterpropagating light waves, and the heating process, associated with the fluctuation of momentum due to atom-light interaction. Since atoms are trapped in the region of the minimum magneto-optical potential, where the magnetic field is zero, the temperature in a MOT
relies mainly on estimates of the temperature of atoms in optical molasses \cite{metcalf}. In particular, in the field, formed by running waves with opposite circular polarizations ($\sigma_+ - \sigma_-$ configuration), the sub-Doppler friction mechanism leads to the possibility of cooling below the Doppler temperature limit \cite{Dalibard89}. 

Note that the semiclassical approach formally allows to analyze the laser cooling in the MOT beyond the optical molasses approximation, since the expressions for the forces on atoms can be obtained, in the general case, with taking into account the presence of a magnetic field \cite{pru2008}. However, the detailed investigation of laser cooling in optical molasses taking into account the quantum recoil effects shows that the quantum approach leads to differences from the results of the semiclassical approach \cite{Kirpichnikova2020}. These quantum recoil effects become important for the laser cooling of atoms with an insufficiently small recoil parameter $\varepsilon_R = \omega_R/\gamma \gtrsim 0.01$ (where $\hbar \omega_R$ is the recoil energy of the atom, and $\gamma$ is the natural linewidth) \cite{Brazhnikov,Wilkowski} and especially for narrow-line optical transitions, where $\varepsilon_R  \gtrsim 0.1$ \cite{pru2019}. Therefore, the similar quantum recoil effects can be important in the MOT as well.

However, within the quantum approaches, which allow for the accurate consideration of quantum recoil effects, numerical analysis of laser cooling in a MOT is extremely difficult. For example, in advanced quantum Monte Carlo approaches \cite{Molmer93,Molmer3D}, the characteristic spatial scale is determined by the choice of minimal partition of the discretization of the wave function in the momentum space, which is typically about $\Delta p /\hbar k \simeq 0.1- 0.01$ due to the limited resources of computing systems. This, as a result of the uncertainty principle, leads to spatial scales of several tens of wavelengths only which is quite enough to describe the laser cooling in optical molasses \cite{pru2007,pru2011,DalibardEPL}.
However, the MOT trapping area is much larger. Therefore, the atomic kinetics in the MOT requires taking into account two various spatial scales $\Lambda_{micro}$ and $\Lambda_{macro}$. The first one is the scale of variation of the light field parameters determined by the wavelength, $\Lambda_{micro} \ll \lambda$. The second macroscopic scale is determined by the region in which the cold atoms are located in the MOT potential, $\Lambda_{macro} \sim 10^4 - 10^5 \lambda$, and may typically reaches a centimeter size. As a result, to describe the kinetics of atoms in a MOT within the quantum Monte Carlo approaches in momentum space one should choose a numerical mesh for the wavefunction calculations on the scale $-p_{max}\,...\,p_{max}$, where $p_{max }= 2\pi\hbar/\Lambda_{micro}$. While the mesh discretization should be considerably fine with the size $\Delta p = 2\pi\hbar/\Lambda_{macro}$ to cover the entire region of the atom localization in the MOT. Thus, the numerical mesh requires extremely large number of the nodes $L_{macro}/L_{micro}\simeq 10^6-10^7$, which, taking into account the real structure of the internal states of atoms, imposes extremely high requirements on the resources of computing systems, making such a task quite challenging for numerical implementation.

In this paper, we propose an effective method, which allows to consider kinetics of atoms in the MOT from first principles in the framework of a quantum approach. Our method is based on the solution of the quantum kinetic equation for the atomic density matrix with full consideration of quantum recoil effects due to interaction with light field. We directly obtain a steady-state solution and also underline some peculiarities in the kinetics of atoms in the MOT.

\section{Quantum kinetic equation for laser cooling of atoms in a MOT}
Let us consider one-dimensional model of the MOT. The monochromatic laser field 
\begin{equation}\label{ffield}
{\bf E}(z, t) = {\mbox Re} \left\{ {\bf E}(z) e^{-i\omega t}\right\} \, ,
\end{equation}
with the local complex vector amplitude  ${\bf E}(z)$, which is formed by counterpropagating waves with opposite circular polarizations
\begin{equation} \label{field}
{\bf E}(z) = E_0 \left({\bf e}_+ e^{ikz} + {\bf e}_- e^{-ikz} \right)  \, .
\end{equation}
It is so-called $\sigma_+$--$\sigma_-$ configuration, where ${\bf e}_{\pm} = \mp \left({\bf e}_x \pm {\bf e}_y \right)/\sqrt{2}$ are unit polarization vectors of the running waves, and  $E_0$ is the scalar amplitude of each running wave. The light frequency $\omega$ is chosen close to resonance of the atomic transition $F_g \to F_e$, where $F_g$ and $F_e$ are the total angular momenta of the ground and excited states, which are described by the Zeeman sub-levels $|F_g,m_g \rangle$ and $|F_e,m_e \rangle$, respectively ($-F_{e,g} \leq m_{e,g} \leq F_{e,g}$).
The spatially nonuniform magnetic field of the trap 
\begin{equation}\label{magnetic}
{\bf B}(z) = B(z) {\bf e}_z
\end{equation}
is chosen along the $z$-axis.

We describe the kinetics of atoms within the framework of the quantum kinetic equation for the atomic density matrix ${\hat \rho}$, which allows us to take into account the quantum recoil effects due to an interaction of atoms with the light field. This equation has the general form
\begin{equation}\label{QQE}
\frac{\partial}{\partial t} {\hat \rho} = -\frac{i}{\hbar} \left[ {\hat H}, {\hat \rho} \right] +{\hat \Gamma}\left\{ {\hat \rho} \right\} \, ,
\end{equation}
where ${\hat H}$ is the Hamiltonian of the system, and the last term ${\hat \Gamma}\left\{ {\hat \rho} \right\}$ describes the relaxation processes in the quantum system. Within the single-particle approximation, i.e. under conditions of low atomic density (so called ``temperature limited'' regime of MOT \cite{Dalib95}), the action of the operator ${\hat \Gamma}\left\{ {\hat \rho} \right\}$ is reduced to the evolution of ${\hat \rho}$ due to spontaneous decays only. The interest in such an analysis is motivated on the one hand by the fact that single-particle approximation allows to determine the limit of laser cooling in MOT, and on the other hand, it serves as a basic step for consideration of high-space-density MOT regimes \cite{Dalib95}.

For density matrix formalism, the most used representations are momentum representation ${\hat \rho}(p_1, p_2)$, coordinate representation ${\hat \rho}(z_1, z_2)$, and Wigner representation ${\hat \rho}(z, p)$. In our approach, we use the coordinate representation, for which the equation (\ref{QQE}) can be written as
\begin{eqnarray}\label{QQEz}
    \frac{\partial}{\partial t} {\hat \rho}(z_1,z_2) &=& -\frac{i}{\hbar} \left[ {\hat H}(z_1) {\hat \rho}(z_1,z_2) - {\hat \rho}(z_1,z_2)  {\hat H}(z_2)\right] \nonumber \\
    &+&{\hat \Gamma}\left\{ {\hat \rho}(z_1,z_2) \right\} \, .
\end{eqnarray}
The Hamiltonian ${\hat H}(z)$ is split into the sum of several contributions
\begin{equation}\label{ham}
    {\hat H}(z) = \frac{{\hat p}^2}{2M}+{\hat H}_0 + {\hat V}_{ED}(z) + {\hat H}_B(z) \,.
\end{equation}
Here ${\hat p} = -i\hbar\, \partial/\partial z$ is the momentum operator, ${\hat H}_0$ is the Hamiltonian of free atom in the rest frame, ${\hat V}_{ ED}$ is the operator of atom-light interaction, and ${\hat H}_B$ describes the interaction of an atom with a magnetic field ${\bf B}(z)$. In the resonant approximation, we can write
\begin{equation}
    {\hat H}_0 = -\hbar \delta {\hat P}_e \, ,
\end{equation}
where $\delta = \omega - \omega_0$ is the detuning of the field frequency $\omega$ from the atomic transition frequency  $\omega_0$, and ${\hat P}_e = \sum_{m_e} |F_e,m_e \rangle \langle F_e, m_e |$ is a projection onto the excited state of the atom. The atom-light interaction in the dipole approximation has the form
\begin{eqnarray}\label{Ved}
    {\hat V}_{ED}(z)&=& {\hat V}(z) + {\hat V}^{\dagger}(z) \, , \nonumber \\
    {\hat V}(z) &=&  - \left({\bf E}(z) \cdot {\bf \hat D} \right){\bar d}/2\hbar \, ,
\end{eqnarray}
where ${\bar d}$ is the reduced dipole moment and ${\bf \hat D}$ is the matrix block of the dipole moment operator, which can be written in a circular basis as
\begin{equation}\label{Dq}
{\hat {\bf D}} = \sum_{s=0,\pm1} {\hat D}_s \,{\bf e}^s \,.
\end{equation}
The circular components of ${\hat {\bf D}}$ in accordance with the Wigner-Eckart theorem \cite{Varshalovich} are expressed in terms of the Clebsch-Gordan coefficients
\begin{equation}
    {\hat D}_s = \sum_{m_e, m_g} C^{F_e,m_e}_{Fg,m_g;\, 1,s} \,|F_e, m_e \rangle \langle F_g, m_g| \,, \, (s=0, \pm1).
\end{equation}
Thus, the interaction operator in the $\sigma_+$--$\sigma_-$ configuration (\ref{field}) takes the form
\begin{equation}\label{v00}
    {\hat V}(z) = \hbar \frac{\Omega}{2} \left({\hat D}_{+1} e^{ikz} + {\hat D}_{-1} e^{-ikz}\right) \, ,
\end{equation}
where $\Omega = -E_0 {\bar d}/\hbar$ is the Rabi frequency per running wave.
The last term in Hamiltonian (\ref{ham}) in the coordinate representation has the form
\begin{equation}
{\hat H}_B(z) = -\big( {\boldsymbol {\hat \mu}} \cdot {\bf B}(z)\big)\, ,
\end{equation}
where ${\boldsymbol {\hat \mu}}$ is the magnetic moment operator.

The operator ${\hat \Gamma}\left\{ {\hat \rho}\right\}$ of the kinetic equation (\ref{QQE}), describing the relaxation of the density matrix due to spontaneous decays with taking into account the recoil effects, is given by
\begin{eqnarray}\label{Gspp}
  &{\hat \Gamma}\left\{ {\hat \rho}\right\}& = \frac{\gamma}{2} \left( {\hat P}_e \,{\hat \rho} + {\hat \rho}\,{\hat P}_e\right)  -  \\
  &-& \gamma \sum_{s=0,\pm 1} \int_{-1}^{+1} {\hat D}^{\dagger}_s\, e^{-iku{\hat z}}{\hat \rho}\,e^{iku{\hat z}}{\hat D}_s \,K_s(u) \, du \, . \nonumber
\end{eqnarray}
The functions $K_0(u) = 3(1-u^2)/4$ and $K_{\pm1}(u) = 3(1+u^2)/8$ are defined by the probability of spontaneous photon emission with polarization $s = 0, \pm 1$ at an angle $\theta$ with respect to the $z$-axis ($u = \cos(\theta)$). 

In the coordinate representation for the variables
\begin{eqnarray}
z &=& (z_1+z_2)/2 \, ,\nonumber \\
q & = & z_1-z_2  \, ,
\end{eqnarray}
the expression (\ref{Gspp}) takes the simple form
\begin{eqnarray}\label{Gspz}
  {\hat \Gamma}\left\{ {\hat \rho}(z,q)\right\} &=& \frac{\gamma}{2} \left( {\hat P}_e\, {\hat \rho}(z,q) + {\hat \rho}(z,q){\hat P}_e\right)  \nonumber \\ 
  &-&\gamma \sum_{s=0,\pm 1} \kappa_s(q)\,  {\hat D}^{\dagger}_s \,{\hat \rho}(z,q)\, {\hat D}_s  \, ,
\end{eqnarray}
where functions $\kappa_s(q)$ 
\begin{eqnarray}
    \kappa_{\pm1} &=& \frac{3}{2}\left( \frac{\sin(kq)}{kq}-\frac{\sin(kq)}{(kq)^3}+\frac{\cos(kq)}{(kq)^2}\right) \, , \nonumber \\
    \kappa_{0} &=& 3\left( \frac{\sin(kq)}{(kq)^3}-\frac{\cos(kq)}{(kq)^2}\right) \, 
\end{eqnarray}
follow from integral part in (\ref{Gspp}).

Note that the solution of the equation (\ref{QQEz}) gives the complete information on the internal and motional states of an atom in the MOT. For example, the distribution function in the phase space ${\cal F}(z,p)$ can be obtained by convolution with respect to the internal states of the density matrix in the Wigner representation
\begin{equation}
    {\cal F}(z,p) = {\mbox{Tr}} \left\{{\hat \rho}(z,p) \right\} \, ,
\end{equation}
while the transformation from the coordinate representation to the Wigner representation is given by 
\begin{equation}\label{Wigner}
   {\hat \rho}(z,p) = \int_{-\infty}^{\infty} {\hat \rho}(z,q)\,\, e^{-i\,q\,p} \, \frac{dq}{2\pi} \, .
\end{equation}
The normalization condition for Wigner density matrix 
\begin{equation}
\int_{-\infty}^{\infty} {\mbox{Tr}} \left\{{\hat \rho}(z,p)\right\} dz\, dp =1
\end{equation}
in coordinate representation has a form
\begin{equation}\label{normintz}
\int_{-\infty}^{\infty} {\mbox{Tr}} \left\{{\hat \rho}(z,q=0)\right\}dz =1\, .
\end{equation}

Without the magnetic field, the method for solving equation (\ref{QQEz}) with full account for quantum recoil effects in an arbitrary one-dimention configuration of the light field was presented in our works \cite{pru2007,pru2011,pru2016}. The presence of an additional contribution ${\hat H}_B$ in the Hamiltonian (\ref{ham}) with an arbitrary coordinate dependence $B(z)$ prevents the use of the previously developed approaches. Below, we present a new method, which allows to solve the equation (\ref{QQEz}) taking into account the contribution from the spatially nonuniform magnetic field (\ref{magnetic}).

\section{Continued fraction method to solve the quantum kinetics equation for atoms in a MOT}
Note that the $\sigma_+ - \sigma_-$ configuration of the light field represents a field with linear polarization in each point, the orientation of which changes by an angle  $\theta = kz$ when moving along the $z$-axis. Therefore, we will use a rotating basis, the transformation to which is given by the operator
\begin{equation}
{\hat U}(z) = \exp{(i k z {\hat F}_z)} \, ,
\end{equation}
where ${\hat F}_z$ is the component of the total angular momentum operator ${\bf \hat F}$ along the $z$-axis. This transformation allows eliminating the fast spatial dependence on the wavelength scales in the atom-light interaction operator (\ref{v00}) and, consequently, in the solution for the transformed atomic density matrix $\hat{ \tilde{\rho}}(z,q)$
\begin{equation}\label{transf}
    \hat{ \tilde{\rho}}(z,q) = {\hat U}^{\dagger}(z) {\hat \rho}(z,q) {\hat U}(z) \, .
\end{equation}
Indeed, the equation (\ref{QQEz}) in the new basis takes the form
\begin{widetext}
\begin{eqnarray}\label{QQErotate}
  \frac{\partial}{\partial t} \hat{ \tilde{\rho}} &-& \frac{i}{M}\frac{\partial}{\partial z}\frac{\partial}{\partial q}\hat{ \tilde{\rho}} = -\frac{k}{M} \frac{\partial}{\partial q} \left[ {\hat F}_z \,  \hat{ \tilde{\rho}} -\hat{ \tilde{\rho}}\, {\hat F}_z \right] -\frac{i}{\hbar} \left[{\hat H}_0\, \hat{ \tilde{\rho}} - \hat{ \tilde{\rho}} \, {\hat H}_0 \right] +{\hat \Gamma}\left\{\hat{ \tilde{\rho}} \right\}  \nonumber \\
  &-& \frac{i}{\hbar}\left[{\hat W}_{ED}(q) \, \hat{ \tilde{\rho}} - \hat{ \tilde{\rho}} \,{\hat W}_{ED}(-q) \right] -\frac{i}{\hbar} \left[{\hat H}_B(z+q/2) \, \hat{ \tilde{\rho}} - \hat{ \tilde{\rho}} \,{\hat H}_B(z-q/2) \right] \, ,
\end{eqnarray}
\end{widetext}
in which $\hat{\tilde{\rho}}(z,q)$ has a smooth dependence on the coordinate determined by the macroscopic change of the magnetic field in the Hamiltonian $\hat{H}_B(z)$. The atom-light interaction operator in the rotating basis is transformed into
\begin{eqnarray}
  {\hat W}_{ED}(q) &=& {\hat U}^{\dagger}(z) \,{\hat V}_{ED}(z+q/2) \,{\hat U}(z)    \\
   &=& \frac{\hbar \Omega}{2} \left( {\hat D}_{+1} e^{i\,k\,q/2}+  {\hat D}_{-1} e^{-i\,k\,q/2} \right) +h.c.\, , \nonumber 
\end{eqnarray}
is a function of variable $q$ only, which characterizes the spatial correlation of the atomic states between two points in the space $z_1 = z+q/2$ and $z_2 = z-q/2$. 

For numerical solution of (\ref{QQErotate}) let us choose a fine mesh for the variable $q$ in the range from $-q_{max}$ to $q_{max}$ in the nodes of which $q$ takes discrete values $q_i$. For the uniform mesh, the derivative $\partial/\partial q$ in equation (\ref{QQErotate}) can be approximated using a finite difference method
\begin{equation}
    \frac{\partial }{\partial q} \hat{ \tilde{\rho}}(z,q) \approx \frac{1}{2\Delta_q} \left( \hat{ \tilde{\rho}}(z,q_{i+1}) - \hat{ \tilde{\rho}}(z,q_{i-1})\right) \, ,
\end{equation}
with $\Delta_q$ as the mesh step. In fact, the parameter $\Delta_q$ determines the maximal momentum of the atoms $p_{max} = \hbar k \,\pi/\Delta_q$ to be accounted for the numerical implementation for the atomic density matrix in the Wigner representation (\ref{Wigner}). In addition, $q_{max}$ defines the minimal width of peaks $\Delta p = \hbar k\,\pi/q_{max}$ in the momentum distribution, which can be correctly described for the given mesh.  Furthermore, we define a mesh for the variable $z$ in the range $-z_{\text{max}} \dots z_{\text{max}}$ with a step $\Delta_z$. In the case of uniform mesh, the operator $\partial /\partial z$ in equation (\ref{QQErotate}) can be  approximated by
\begin{equation}\label{ddz}
\frac{\partial }{\partial z} \hat{ \tilde{\rho}}(z,q_i) \approx \frac{1}{2\Delta_z} \left( \hat{ \tilde{\rho}}(z_{n+1},q_{i}) - \hat{ \tilde{\rho}}(z_{n-1},q_{i})\right) \, . 
\end{equation}

\begin{figure}[t]
\centerline{\includegraphics[width=3.3 in]{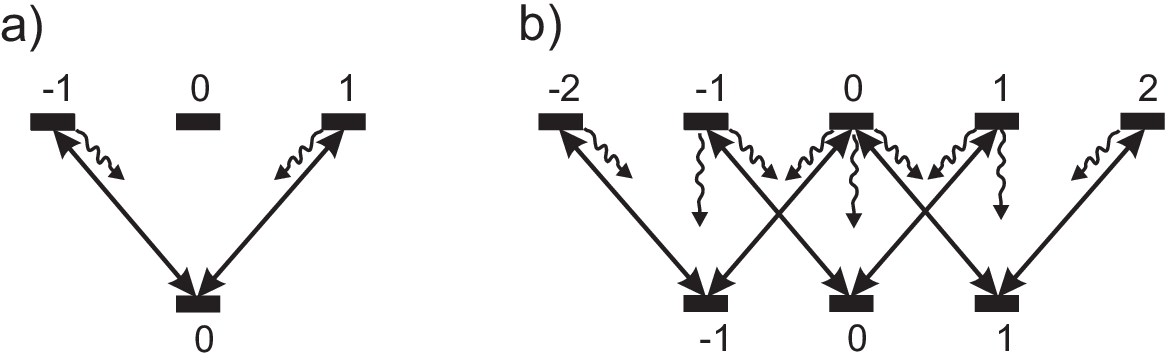}}
\caption{The scheme of Zeeman sublevels interacting with the light field for atoms with optical transitions $F_g=0 \to F_e=1$ (a) and $F_g=1 \to F_e=2$ (b). The double arrows indicate the light induced transitions, while the wavy arrows indicate transitions due to spontaneous decay.} \label{fig:F1}
\end{figure}

Next, we turn from the matrix notation for the density matrix $\hat{ \tilde{\rho}}(q_i,z_n)$ to the vector form $\vec{\bf \rho}$, in which the matrix elements $\hat{ \tilde{\rho}}_{m,m'}(q_i,z_n)$ are written sequentially as a column vector with indices ${\rho}_{m,m',q_i,z_n}$. It is so called the Liouville space representation of the density matrix. Here, $m$ and $m'$ are indexes that describe all Zeeman sublevels of the atom (see Fig.\ref{fig:F1}). In this representation, the equation for the density matrix (\ref{QQErotate}) can be written in general form
\begin{equation}\label{liuv0}
    \frac{\partial }{\partial t}\vec{\bf \rho} = \hat{\cal L}\, \vec{\bf \rho} \, ,
\end{equation}
where the Liouvillean $\hat{\cal L}$ is the matrix of size ${\cal N}\times {\cal N}$, with ${\cal N} = 4(F_e+F_g+1)^2 \, N_z \, N_q$. Here, $N_z$ and $N_q$ are the number of nodes in the meshes for variables $z$ and $q$. The steady-state solution of (\ref{liuv0}) satisfies the equation
\begin{equation}\label{stt}
    \hat{\cal L}\, \vec{\bf \rho} = 0 \, ,
\end{equation}
which must be supplemented by the normalization condition (\ref{normintz}) that can be rewritten in the form
\begin{equation}\label{norm0}
{\mbox Tr}\left\{\hat{ \tilde{\rho}}(z=0,q=0) \right\} = const   \, , 
\end{equation}
since the determinant of the Liouvillian $\hat{\cal L}$ is equal to zero. 

\begin{figure*}[t]
\centerline{\includegraphics[width=6.0 in]{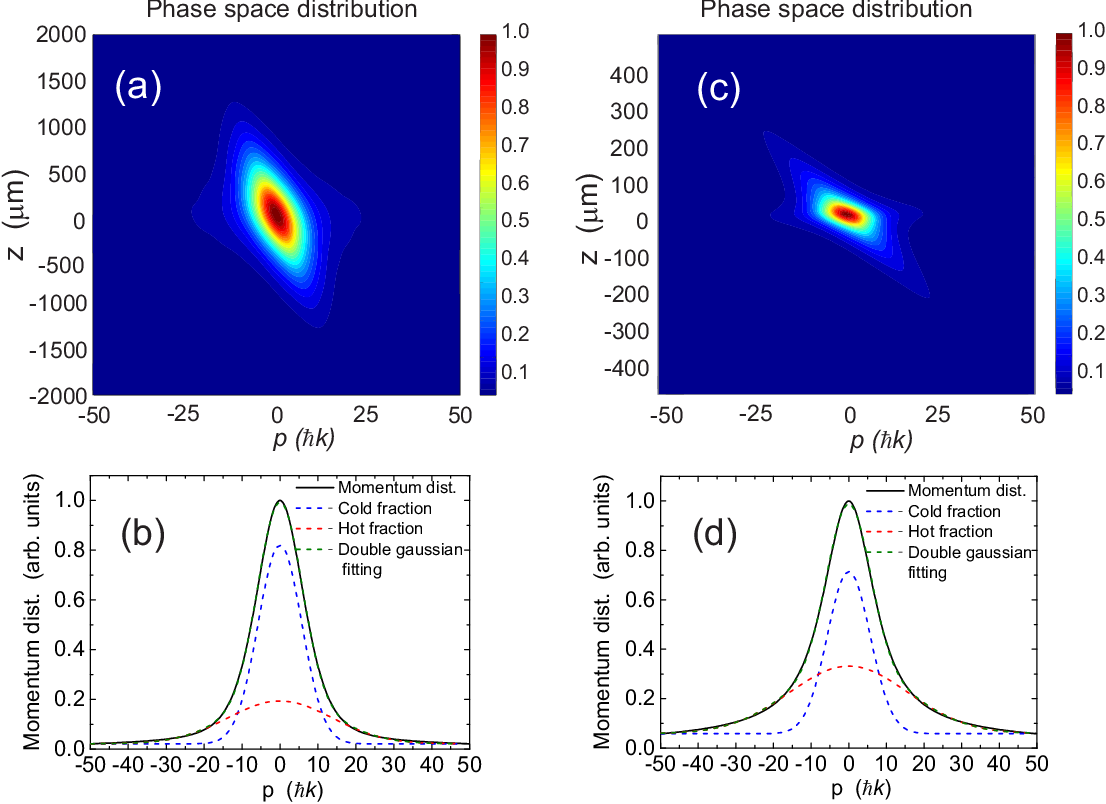}}
\caption{The steady-state solution of the quantum kinetic equation (\ref{QQErotate}) in a MOT for  $^{87}$Rb atoms with optical transition $F_g = 2 \to F_e = 3$, the recoil parameter $\varepsilon_R = \hbar k^2/(2M\gamma) = 6 \cdot 10^{-4}$. (a,c) Wigner phase space distribution function ${\cal F}(z,p)$. (b,d) Momentum distribution ${\cal F}(p) = \int {\cal F}(z,p)\, dz$ and its fitting by double Gaussian functions, separating into ``hot'' and ``cold'' fractions. The magnetic field gradient $\beta = 1$ G/cm (a,b) and $\beta = 10$ G/cm (c,d). The Rabi frequency of running waves $\Omega = \gamma$, the detuning $\delta = -5 \gamma$.} \label{fig:F2}
\end{figure*}

We emphasize that the dimensionality of the Liouville space $\hat{\cal L}$ is sufficiently huge. To reduce the dimensionality, we first note that for considered schemes of atom-light interaction (see Fig.\ref{fig:F1}), the number of non-zero matrix elements $N_{int}$, describing the internal states of the atom, is less than $4(F_e+F_g+1)^2$. For example, for the optical transition $F_g=0 \to F_e=1$ among the total number of 16 matrix elements there only $N_{int} = 9$ are non-zero that describe the changes in the internal states of the density matrix. For the optical transition $F_g=1 \to F_e=2$, among the total number of 64 matrix elements, only $N_{int} = 34$ are non-zero. Thus, the overall dimensionality of the matrix $\hat{\cal L}$ can be reduced to ${\cal N}'\times {\cal N}'$, with ${\cal N}'=N_{int}\, N_q \, N_z$, and still remains large enough, significantly complicating the numerical solution of equation $(\ref{liuv0})$. 

In the following, we develop the effective method, which allows to obtain a steady-state solution $(\ref{stt})$ by using Liouvillean matrices of reduced dimension ${\cal N}_r\times {\cal N}_r$, where ${\cal N}_r=N_{int}\, N_q$ only. This method is based on continued fractions algorithm. For qualitative understanding the computational problem we note, that even for $N_z=100$ the introduced method allows to save up the computational memory by 10,000 times. Moreover, in some cases, as will be shown below for a MOT with a large magnetic field gradient, the number of spatial points must be significantly increased for a correct description of the trap. In our calculations, this parameter can reach $N_z=500$.

The essence of our method is to construct a set of vectors $\vec{\bf \rho}_n$ of ${\cal N}_r$ elements for each point in the space $z_n$, corresponding to the Liouville representation of the density matrix $\hat{\tilde{\rho}}(q_i,z_n)$. The definition (\ref{ddz}) defines a finite-difference scheme for the equation (\ref{QQErotate}) in the reduced Liouville space
\begin{equation}\label{dynamic}
    \frac{\partial }{\partial t} \vec{\bf \rho}_n -{\hat R}\,\vec{\bf \rho}_{n+1} +  {\hat R}\,\vec{\bf \rho}_{n-1} = {\hat L}_n \, \vec{\bf \rho}_n  \, ,
\end{equation}
where the reduced Liouvillian operator ${\hat L}_n$ is defined by the right-hand side of the equation (\ref{QQErotate}) at the point $z_n$, while 
\begin{equation}
  {\hat R}= \frac{i}{2M\Delta_z} \frac{\partial}{\partial q} \, , 
\end{equation}
corresponds to the operator of the second term in the left-hand side of Eq.\,(\ref{QQErotate}). Consequently, the steady-state solution of (\ref{dynamic}) satisfies the following recursive relation
\begin{equation} \label{CF}
    -{\hat R}\,\vec{\bf \rho}_{n-1} +{\hat L}_n \, \vec{\bf \rho}_n +{\hat R}\,\vec{\bf \rho}_{n+1} = 0 \, ,
\end{equation}
which can be solved by the matrix continued fractions algorithm. 

Let us choose the point $z_0$ near the center of the trap. We assume the indices $n > 0$ correspond to positions $z_n > z_0$, and indices $n<0$ correspond to $z_n<z_0$. For ideces  $n>0$ we define matrices ${\hat S}_n$ as follows
\begin{equation}\label{s}
  \vec{\bf \rho}_{n} = {\hat S}_n \, \vec{\bf \rho}_{n-1} \, .
\end{equation}
Based on the Eq.\,(\ref{CF}), such matrices satisfy the equation
\begin{equation}
\left(-{\hat R} + {\hat L}_n\, {\hat S}_n + {\hat R}\,{\hat S}_{n+1} \,{\hat S}_n \right)\vec{\bf \rho}_{n-1} = 0 \, ,
\end{equation}
which leads to the recurrence relation for ${\hat S}_n$
\begin{equation}\label{S}
{\hat S}_n = \left({\hat R}\,{\hat S}_{n+1} + {\hat L}_n  \right)^{-1}{\hat R} \, .
\end{equation}
Similarly, for $n<0$ we define the sequence of matrices ${\hat T}_n$ 
\begin{equation}\label{t}
  \vec{\bf \rho}_{n} = {\hat T}_n \,\vec{\bf \rho}_{n+1} \, ,
\end{equation}
which are satisfying the recurrence relation
\begin{equation}\label{T}
{\hat T}_n = \left({\hat R}\,{\hat T}_{n-1} - {\hat L}_n \right)^{-1}{\hat R} \, .
\end{equation}

\begin{figure}[t]
\centerline{\includegraphics[width=3.0 in]{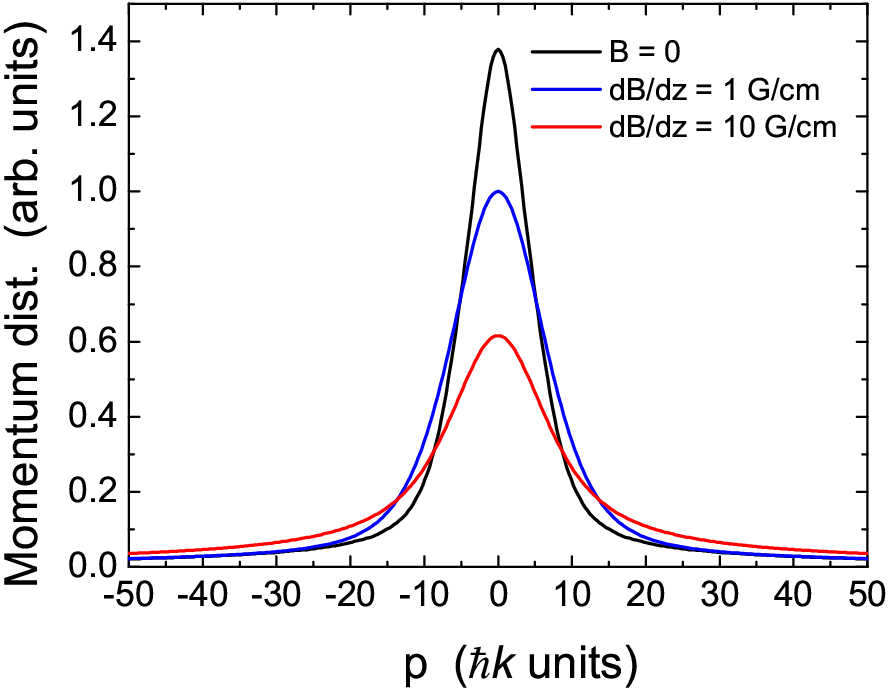}}
\caption{The steady-state momentum distribution of $^{87}$Rb atoms (the optical transition $F_g = 2 \to F_e = 3$) laser cooled in optical molasses for $B=0$ (black line), and cooled in MOT with different magnetic field gradients $\beta = 1$\,G/cm (blue line),  and $\beta = 10$\,G/cm (red line). 
The Rabi frequency of running waves $\Omega = \gamma$, and the detuning $\delta = -5 \gamma$.} \label{fig:F3}
\end{figure}

Next, we assume the atoms are localized near the center of the MOT and the density of atoms decreases at some distance from the trap center. This allows us to define the enough large  $N>0$ for which $\vec{\bf \rho}_{\pm N}= \vec{\bf \rho}_{\pm (N+1)}= 0$, that result to we may choose ${\hat S}_N = {\hat T}_{-N} = I$ is identity matrices. Thus, applying the recursive relations (\ref{S}) and (\ref{T}), we can find all set of matrices ${\hat S}_n$ and ${\hat T}_{n}$ for $|n|<N$. The matrices ${\hat S}_{+1}$ and ${\hat T}_{-1}$ allow us to obtain the equation for $\vec{\bf \rho}_0$ at the central point $z_0$
\begin{equation}\label{eq}
    \big( {\hat R}\,{\hat S}_{+1} +{\hat L}_0 -{\hat R}\,{\hat T}_{-1} \big) \vec{\rho}_0 = 0 \, .
\end{equation}
This equation should be supplemented with the normalization condition (\ref{norm0}), because the determinant of the expression in parentheses in Eq.\,(\ref{eq}) is equal to zero. The solutions for $\vec{\bf \rho}_{n}$ at the points $z_n$ for $|n|>1$ can recursively be obtained from definitions (\ref{s}) and (\ref{t}). The main advantage of the presented method is based, firs of all, on the fact that the matrices ${\hat L}_n$ $\hat{S}_n$, $\hat{T}_n$ and ${\hat R}$ have a significantly reduced dimension compared to the Liouvillian space of the initial Eq.\,(\ref{liuv0}). The second, the choice of $z_{max}$ and the mesh size $\Delta_z$ are adjustable parameters, which determine only the necessary number $N$ of recursion steps in (\ref{S}) and (\ref{T}).

\section{Results of numerical simulations}
For the numerical analysis, we suppose a linear dependence of the magnetic field on coordinate, $B(z) = \beta \, z$ (with $\beta= \partial B/\partial z$ is magnetic field gradient), and the point $z_0 = 0$ corresponds to the trap center with $B=0$. The Fig.\,\ref{fig:F2} demonstrates the steady-state solution of the quantum kinetic equation (\ref{QQErotate}) for  $^{87}$Rb atoms in a MOT formed with the use of D$_2$ line optical transition $F_g = 2 \to F_e = 3$. In particular, the steady-state momentum distribution of the atoms in the MOT ${\cal F}(p) = \int {\cal F}(z,p)\, dz$ (see Figs.\,\ref{fig:F2}(b,d)) is not equilibrium and in general cannot be described in terms of temperature. However, this momentum distribution can be well approximated by the sum of two Gaussian functions, allowing us to separate it into combination of ``cold'' and ``hot'' fractions. For the MOT with magnetic field gradient $\beta = 1$ G/cm, we obtain $T_{cold} \simeq 14.4 \,\mu K$ (with atoms fraction $N_{cold} \simeq 0.7$), which corresponds to sub-Doppler temperature, and $T_{hot} \simeq 260 \mu K$ (with atoms fraction $N_{hot} \simeq 0.3$), which corresponds to temperature on the order of the Doppler limit. The values for $N_{cold}$ and $N_{hot}$ fractions  we calculated from the relative integrated areas of the fitted Gaussians functions. For the MOT with the magnetic field gradient $\beta = 10$ G/cm, we have $T_{cold} \simeq 16 \,\mu K$ ($N_{cold} \simeq 0.5$) and $T_{hot} \simeq 280 \, \mu K$ ($N_{hot} \simeq 0.5$). The spatial distribution ${\cal F}(z) = \int {\cal F}(z,p)\, dp$ of the atoms in the MOT is shown in Figs.\,\ref{fig:F2}(c,f).

The Fig.\,\ref{fig:F3} demonstrates the comparison of the laser cooling in optical molasses (i.e., at zero magnetic field) and in the MOTs with different magnetic field gradients. 
In the optical molasses, two fractions can also be distinguished: a ``cold'' fraction of atoms with $T_{cold} \simeq 7.3 \, \mu K$ ($N_{cold} \simeq 0.7$) and a ``hot'' fraction with $T_{hot} \simeq 230 \, \mu K$ ($N_{hot} \simeq 0.3$). As can be seen, the temperature in MOT is higher than in optical molasses even for single-particle approximation, and tends to the molasses temperature only in the limit of zero magnetic field. Moreover, in the general case, the increment of magnetic field gradient leads to increment of the trapped atoms temperature. These results indicate that the molasses approximation for the MOT does not describe the trap well, even for the ``temperature limited'' regime of the the low-density MOT \cite{Dalib95} with neglecting interatomic interaction.

\begin{figure}[t]
\centerline{\includegraphics[width=3.3 in]{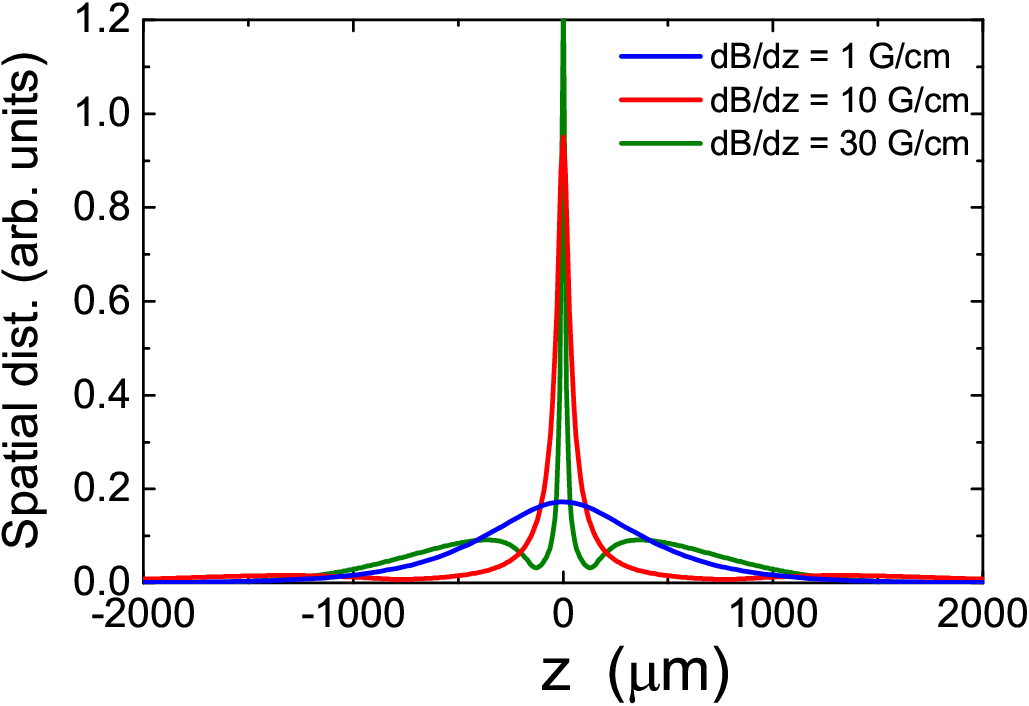}}
\caption{The spatial distribution ${\cal F}(z) = \int {\cal F}(z,p)\, dp$ of atoms in the MOT for $^{87}$Rb atoms (the optical transition $F_g = 2 \to F_e = 3$) for MOT magnetic field gradient of $\beta = 1$ G/cm (blue line),  $\beta = 10$ G/cm (red line), and $\beta = 30$ G/cm (green line). The Rabi frequency of running waves $\Omega = \gamma$, and the detuning $\delta = -5 \gamma$.} \label{fig:F4}
\end{figure}

\begin{figure}[h]
\centerline{\includegraphics[width=3.3 in]{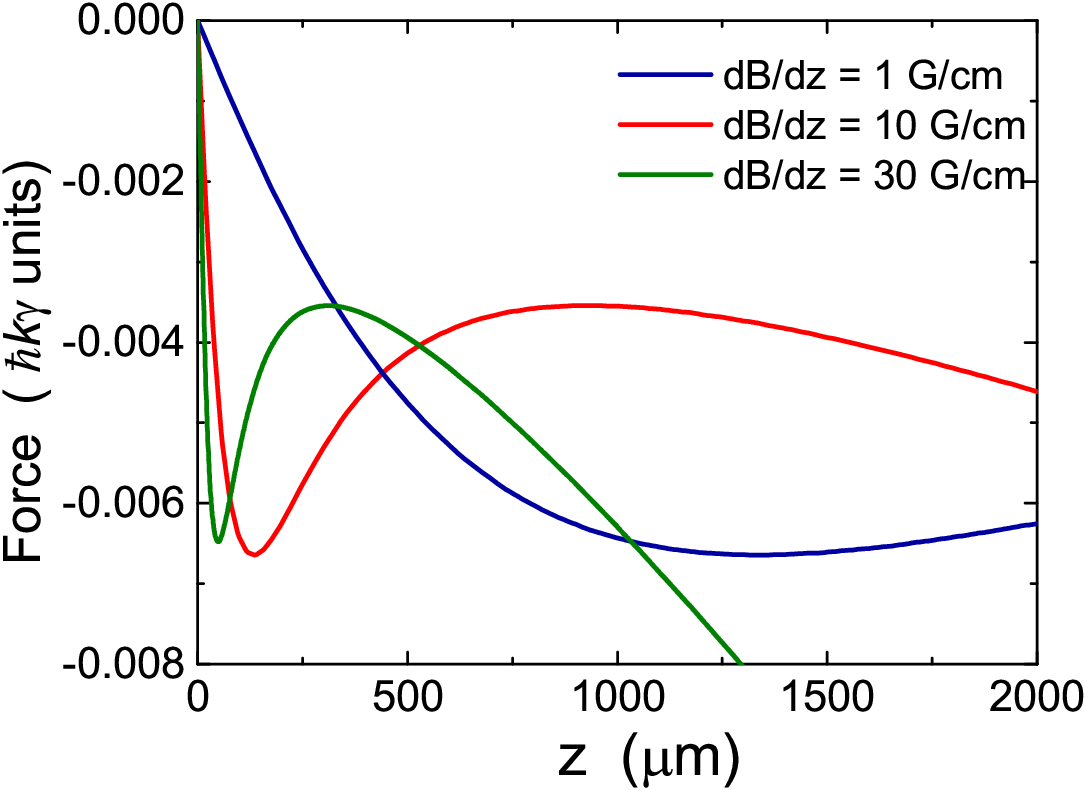}}
\caption{The magneto-optical force forming the MOT trapping potential for $^{87}$Rb atoms (the optical transition $F_g = 2 \to F_e = 3$) for magnetic field gradient of $\beta = 1$ G/cm (blue line),  $\beta = 10$ G/cm (red line), and  $\beta = 30$ G/cm (green line). The Rabi frequency of running waves $\Omega = \gamma$, and the detuning $\delta = -5 \gamma$.} \label{fig:F5}
\end{figure}

In addition, in a MOT with large gradient of the magnetic field, a two-component spatial distribution of atoms can be seen in Fig.\,\ref{fig:F4}. This effect looks very similar to ``two-component'' regime of the MOT described in \cite{Dalib95}, where such an effect is explained for the MOT with a high number of trapped atoms, when the interatomic interaction becomes important: ``As more and more atoms are loaded into the trap the cloud becomes fatter and fatter, until eventually it fills the central, strongly confining region. If further atoms are added the cloud then spills over into the weakly confining surrounding
volume, spreading out to a much larger radius.'' (see in \cite{Dalib95}). However, for the MOT with large magnetic field gradient, we observe such an effect even for the single-particle approximation (see Fig.\,\ref{fig:F4}). For the MOT with magnetic field gradient $\beta = 10$ G/cm there is an increase in the space distribution ${\cal F}(z)$ at the edges of the trap in the region $z \simeq 600 - 2000 \, \mu m$, and for $\beta = 30$\,G/cm in the region $z \simeq 150 - 900 \, \mu m$.  These regions coincide with the regions of decreasing of the restoring force on trapped atom (see Fig.\,\ref{fig:F5}). Thus, the MOT may has such a ``two-component'' spatial regime not only due to increasing trapped atom cloud size by increasing the number of atoms, but also in low-density MOT condition.

\section{Conclusion}
We presented the quantum theory of a MOT developed from first principles and based on the quantum kinetic equation for the atomic density matrix, taking into account quantum recoil effects due to interaction of atom with the light waves. We developed the effective method for solving the quantum kinetic equation within the low-density regime of the MOT (the so-called ``temperature limited'' regime of the MOT \cite{Dalib95}). The importance in analyzing of this regime is related to the fact that it provides a qualitative understanding of the MOT operation and can be used as a basic step for further study of MOT regimes with high density of atoms. This requires a corresponding modification of both the Hamiltonian of the system and the relaxation operator ${\hat \Gamma}\left\{ {\hat \rho} \right\}$ in Eq.\,(\ref{QQE}). As well, the gravity can also be taken into consideration  by adding additional term into the Hamiltonian (\ref{ham}).

Our quantum analysis of the MOT demonstrates several important results. In particular, it was shown that the steady-state of atoms in the MOT has a significantly non-equilibrium nature and, in general, cannot be described by a Gaussian function for the momentum distribution. In fact, two fractions of trapped atoms can be distinguished with Doppler and sub-Doppler temperatures. Moreover, the influence of the magnetic field on the kinetics of atoms in the region of their localization (i.e. near the minimum of the magneto-optical potential) leads to significant differences in the temperature of cold atoms in comparison with optical molasses regime in zero magnetic field. Thus, even the ``temperature limited'' regime of the MOT demonstrates the significant dependence of the temperature on the magnetic field gradient. In addition, a two-component spatial distribution of atoms is observed in the MOT with large magnetic field gradient, even for single-particle approximation. 

Thus, our results demonstrate that the simple models of MOT, which are based on optical molasses approximation with the use of equipartition theorem (for the mean kinetic and potential energies of atoms in a trap) to estimate the temperature and atom cloud size, require corrections.

\section{ACKNOWLEDGMENTS}
O.N.P. deeply thanks L\H{o}rinc S\'{a}rk\'{a}ny for fruitful discussion and interest in this work.

\nocite{*}



\bibliography{quantum_MOT}

\end{document}